\documentclass[a4paper,10pt,fleqn]{article}
\usepackage{epsfig}
\usepackage{graphics,graphicx, subfig}
\usepackage{amsmath}
\usepackage{amssymb}
\usepackage{color}
\usepackage{here}
\usepackage{changepage}
\chardef\at=`\@
\DeclareRobustCommand{\qed}{%
  \ifmmode 
  \else \leavevmode\unskip\penalty9999 \hbox{}\nobreak\hfill
  \fi
  \quad\hbox{\qedsymbol}}

\newcommand{\ep}{\mathbb P}

\begin{document}

\begin{center}
{\huge \bf Stylized Facts and Simulating Long Range Financial
  Data}\footnote[1]{The research has been supported by the Collaborative Research Center ``Statistical modeling of nonlinear dynamic processes'' (SFB 823) of the German Research Foundation, which is gratefully acknowledged.
  }\\
\quad\\
\textbf{Laurie Davies}\\
 Fakult\"at Mathematik, Universit\"at Duisburg-Essen, Germany \\
 e-mail: laurie.davies@uni-due.de\\
 and\\
\textbf{Walter Kr\"amer} \\
 Fakult\"at Statistik, Technische Universit\"at Dortmund, Germany\\
 Phone: 0231/755-3125, Fax: 0231/755-5284 \\ e-mail: walterk@statistik.tu-dortmund.de
\end{center}


\vspace{1cm}
\begin{adjustwidth}{+1cm}{+1cm}
\section*{Abstract}
We propose a new method (implemented in an R-program) to simulate
long-range daily stock-price data. The program reproduces various
stylized facts much better than various parametric models from the
extended GARCH-family. In particular, the empirically observed changes
in unconditional variance are truthfully mirrored in the simulated
data.
\end{adjustwidth}
\vspace{1cm}


\section{Introduction and motivation}

There is considerable interest in empirical finance in generating
daily stock price data which mimic actual stock price behaviour as
closely as possible. Such artificial data are useful, for instance, in
backtesting models for value at risk or in evaluating trading
strategies. The form of mimicking we shall be interested is the the
ability of the model to reproduce certain stylized facts about
financial assets in a quantitative sense. The
concept of stylized facts was introduced in \cite{KAL57}. There have
been several papers on the application of the concept to
financial data; \cite{RYTEAB98}, \cite{CON01}, \cite{HOM02},
\cite{LUSCH05}, \cite{BULBUL06}, \cite{MAMTER10}, \cite{TERZHA11}. These
papers are all dynamic in that they can be used for simulations
once the parameters have been estimated. In general this will require
a small number of parameters as models with a large number of
parameters run into estimation problems. An approach involving some
form of nonparametric estimation cannot be used for simulations unless
the nonparametric component can be adequately randomized. This is the
approach to be taken below. The paper builds on Davies et al. (2012), who
consider daily Standard and Poor's (S+P) $500$ returns over 80
years. The squared returns were approximated by a piecewise constant
function. This can be regarded as a nonparametric approach but in this
paper we model a finer version of the piecewise constant function as a
stochastic process which can then be used to simulate data.

\begin{figure}[!h]
\centering
\includegraphics[width=4cm,height=12cm,angle=-90]{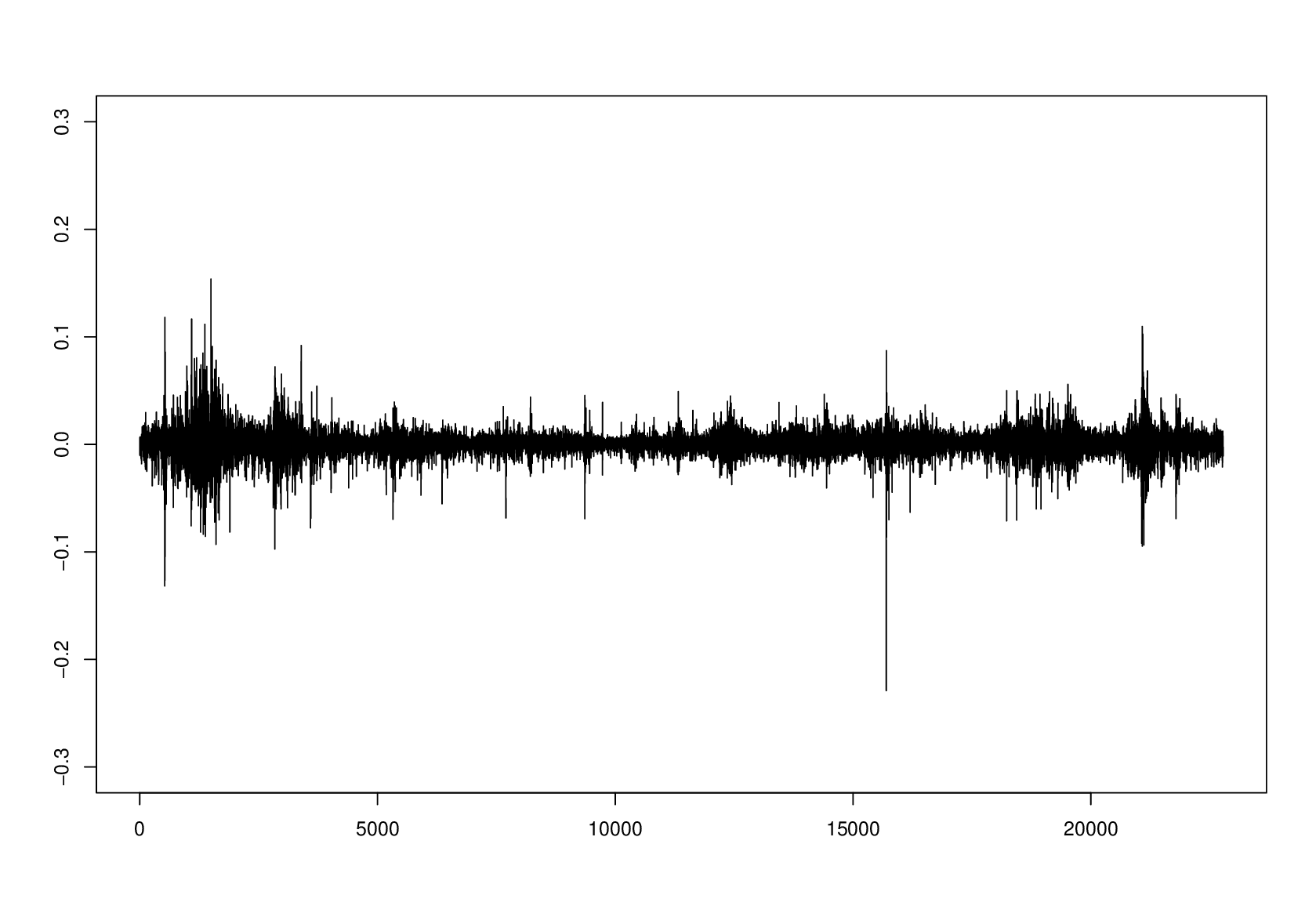}
\includegraphics[width=4cm,height=12cm,angle=-90]{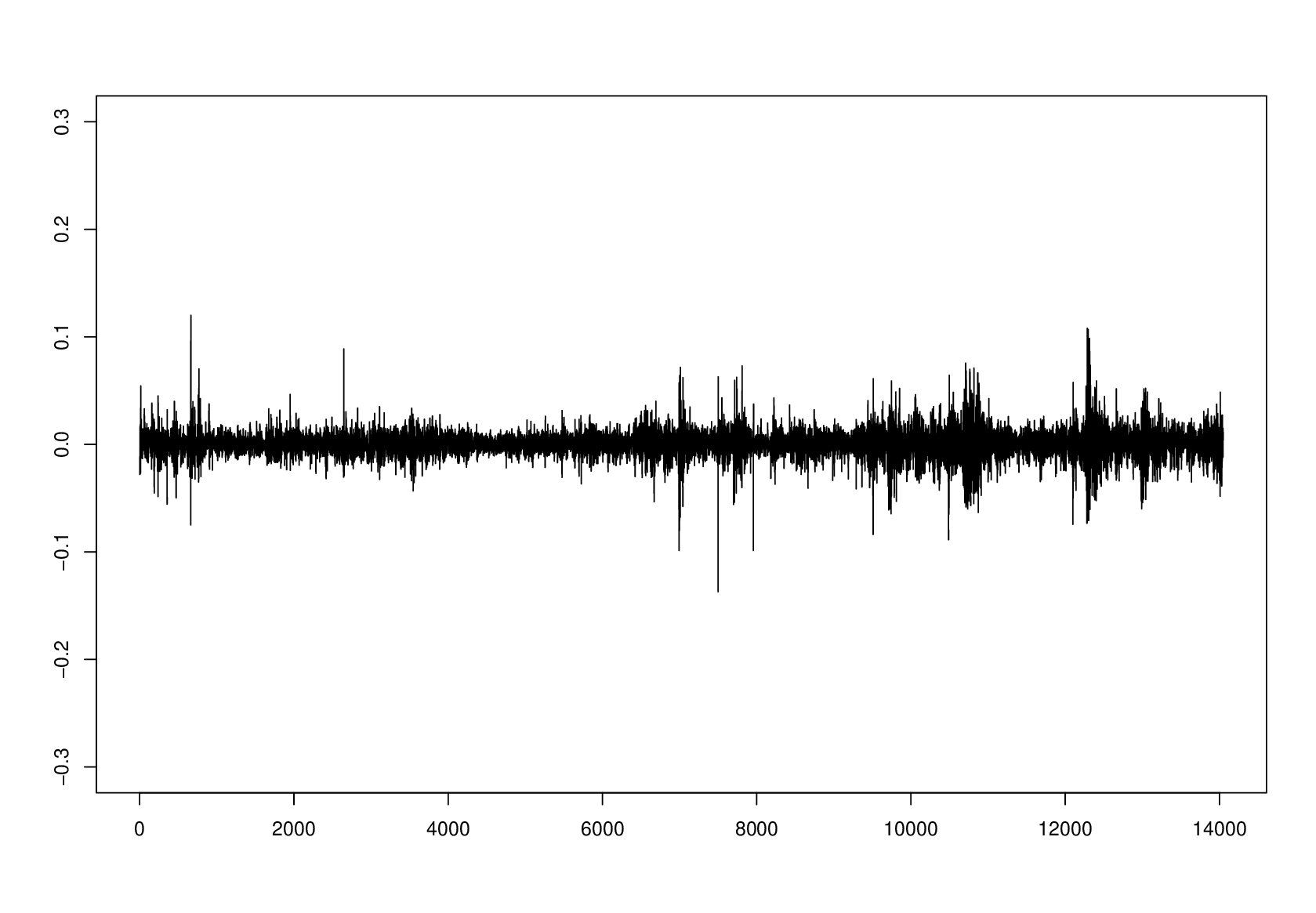}
\caption{Upper panel: The daily returns of S+P 500
  with the zeros removed. Lower panel: the same for DAX.}
\label{fig:SP500_DAX}
\end{figure}

Our main running example is the Standard and Poor's (S+P) $500$ shown
in the upper panel of Figure~\ref{fig:SP500_DAX}. The data consist of
22381 daily S+P returns with the zeros removed. The final day is 24th
July 2015. The second running example is the German DAX index from 30th September 1959 to
19th October 2015 shown in the lower panel of
Figure~\ref{fig:SP500_DAX}. There are 14049 observation of which 14026
are non-zero.
A third set of data sets we shall use are the 30 firms represented in
the German DAX index. The returns are from 1st January 1973, or from
the date the firm was first included in the index, to 13th July 2015.

The question as to whether a model satisfactorily reproduces a
quantified stylized fact or indeed any other quantified property of
the data is typically answered by comparing the empirical value of a
statistic with its value under the model. This was done in
\cite{STUA03} for the unconditional variance using the S+P 500 from
March 4, 1957 to October 9, 2003 excluding the week starting October
19, 1987. The conclusion was that the GARCH(1,1) unconditional
variance was larger than the empirical variance. For the  Standard and
Poor's (S+P) $500$ data at our disposal the unconditional variance is
0.000135 after eliminating zero values. The maximum
likelihood estimates of a GARCH(1,1) model are
\[{\hat \alpha}_0=8.32\text{e-07},\, {\hat \alpha}_1 =0.9106,\, {\hat \beta}_1=
0.08543\]
so that the unconditional second moment under the model is
\[{\hat \sigma}^2={\hat \alpha}_0/(1-{\hat \alpha}_1-{\hat \beta}_1)=
0.000207\]
which is `considerably' larger. This however ignores the variability
of the second moment in simulations. On the basis of 1000 simulation
the 0.05 and 0.95 quantiles of the second moment under the model are
0.000130 and 0.000345 respectively. The empirical value lies between
and has an estimated $p$-value of 0.079 which, while small, would not
be classified as statistically significant.

The same applies to the autocorrelation function. The upper panel of
Figure \ref{fig:acf} shows the ACF for the first 1500 lags for the
absolute S+P 500 values in black: the grey line shows the mean of the
1000 simulations for the GARCH(1,1) model with maximum likelihood
parameters. The lower panel shows nine of the 100 simulations. The
large variability of the ACF values implies that comparing the
empirical values with the means of simulated values can be misleading.
\begin{figure}[!h]
\centering
\includegraphics[width=4cm,height=12cm,angle=-90]{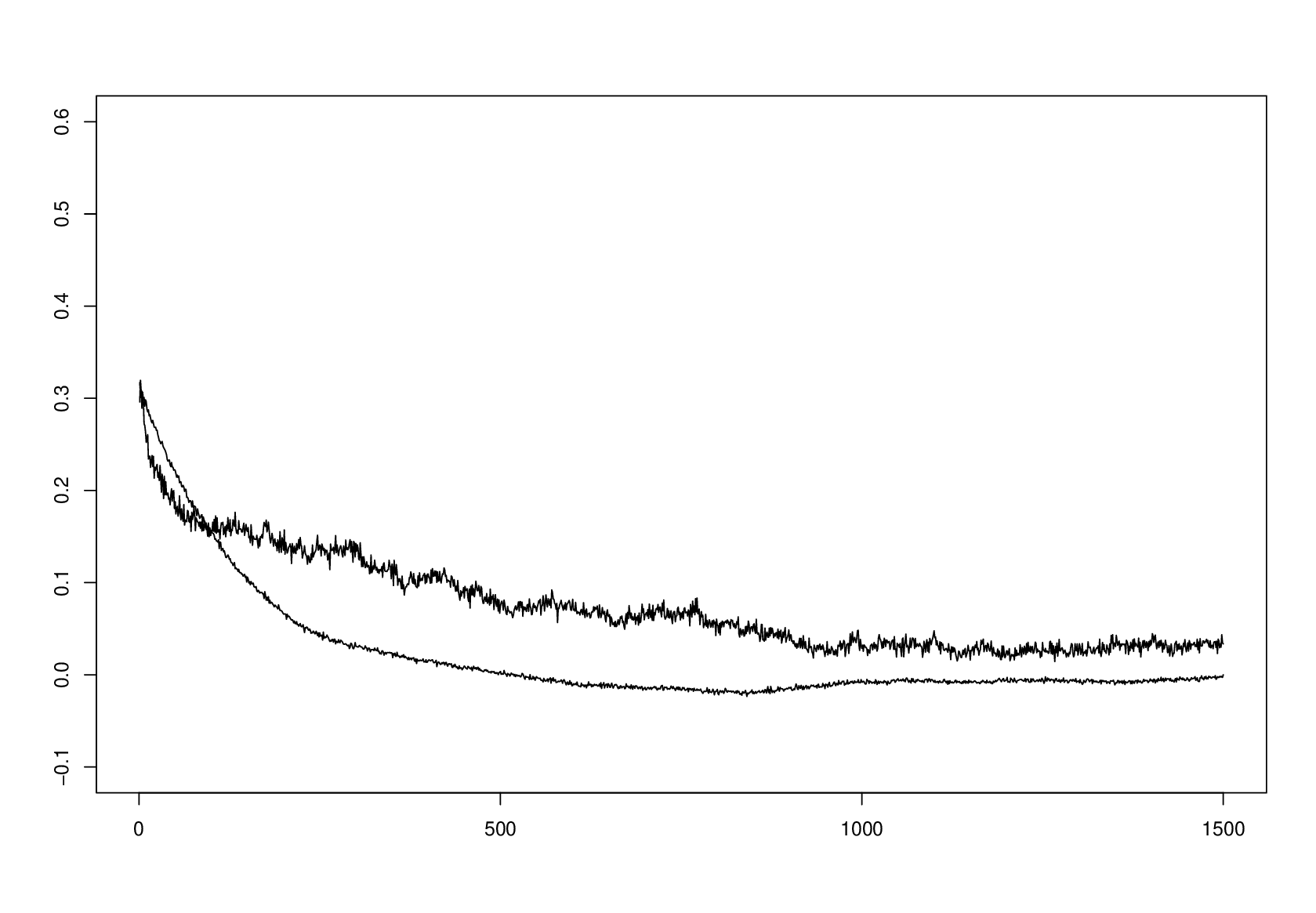}
\includegraphics[width=4cm,height=12cm,angle=-90]{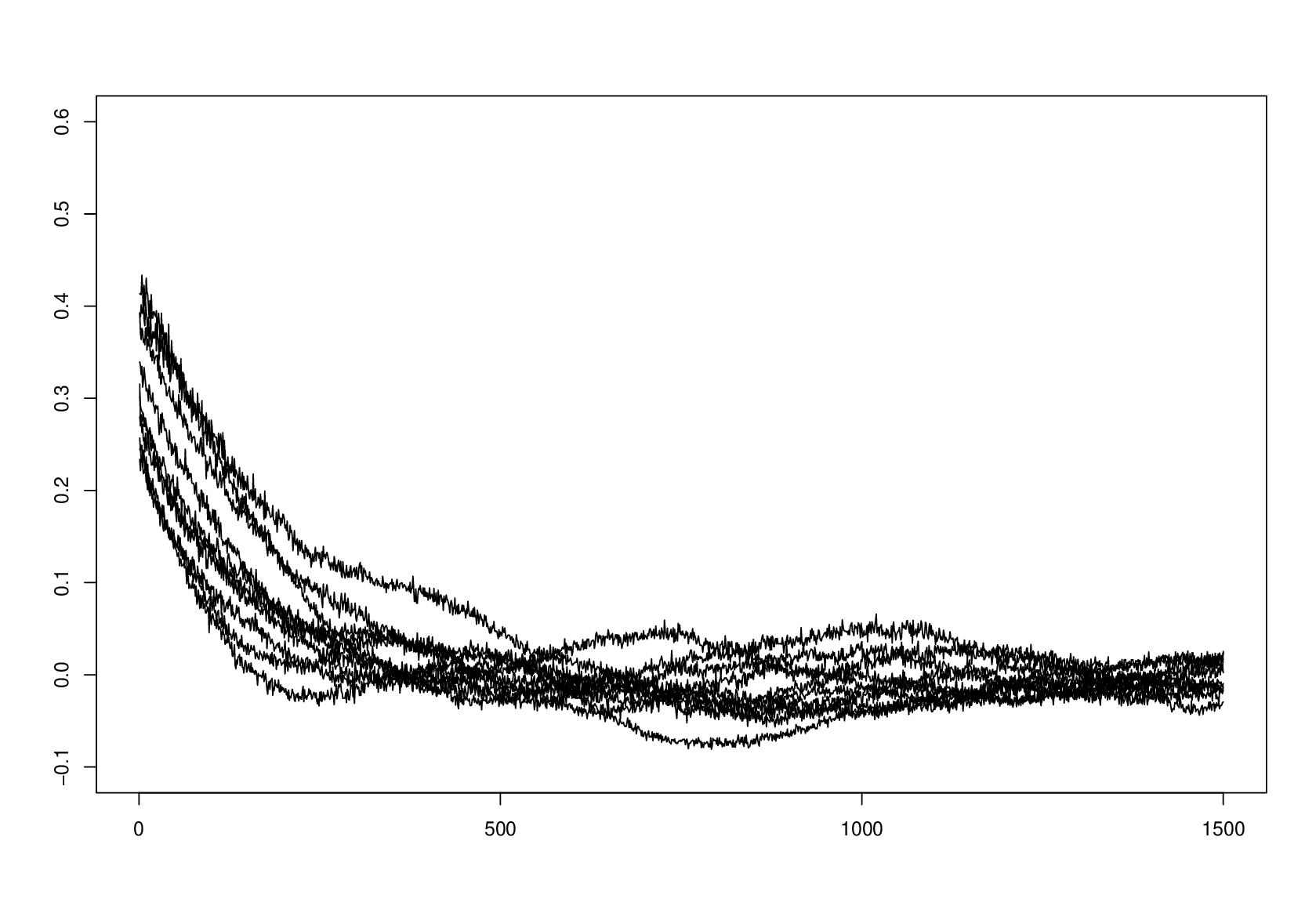}
\caption{Upper panel: the first 1500 values of the empirical ACF of the absolute
  values of the S+P 500 data and the mean ACF of 10 GARCH simulations
  based on the maximum likelihood estimator for the S+P 500. Lower
  panel: the ACFs of the ten simulations \label{fig:acf}}
\label{fig:acf}
\end{figure}

\begin{figure}[!h]
\centering
\subfloat[Panel a: Simulated returns]{
\includegraphics[width=4cm,height=12cm,angle=-90]{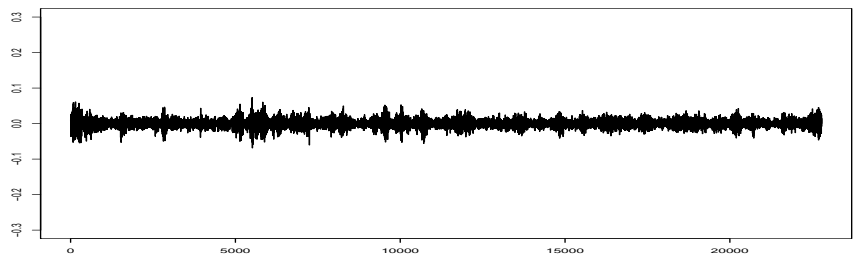}
}

\subfloat[Panel b: Simulated returns for the first 2000 values]{
\includegraphics[width=4cm,height=12cm,angle=-90]{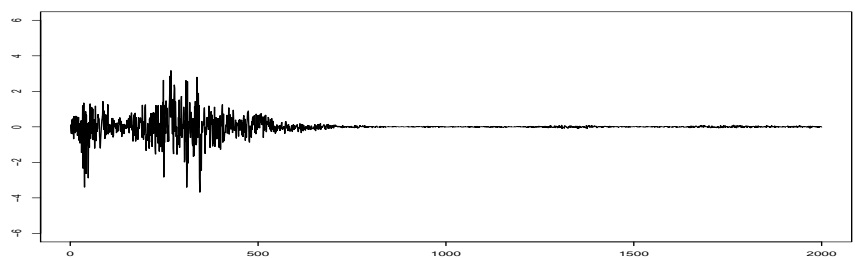}
}
\caption{(a) simulated daily returns from a GARCH(1,1) model
  using the maximum likelihood parameter values based on the complete
  S+P 500 data set. (b) the same for the first 2000 values.}
\label{fig:stpr_garch}
\end{figure}

In \cite{STUA03} the quantitative comparisons mentioned above were
augmented by visual ones. The author compared 24 data sets
generated under the model with the real data (Figure~5.1 of
\cite{STUA03}) and stated `The aspect of the real data is different
from that of the simulated samples'. In this spirit Panel (a) of
Figure~\ref{fig:stpr_garch} shows data simulated under the GARCH(1,1)
model using the maximum likelihood parameters based on the whole S+P
500 data set. Panel (b) shows a simulation for the first 2000 values
based on the maximum likelihood estimators for these values. The
simulated data sets can be compared with the real data shown in
Figure~\ref{fig:SP500_DAX}. The discrepancy visible in Panel (b) is very
large: the average squared return is 0.398  against 0.000197 for the
S+P 500 data. This is due to the fact that the maximum  likelihood
estimates of $\hat{\alpha}_1$ and $\hat{\beta}_1$ in the GARCH(1,1)
model sum to 1.0069 so that the model is not stationary.

Such visual comparisons, also known as `eyeballing', are often used
(see for example \cite{NEYSCOSHA53}, \cite{NEYSCOSHA54}, \cite{DAV95},
\cite{DAV08}, \cite{BUJETAL09}, \cite{DAV14}). Although very useful
and to be recommended they have their limitations. Where
possible the observed differences should be given numerical
expressions and the empirical and simulated values compared. This will
be done in the context of financial series in the remainder of the
paper.

\begin{figure}[!h]
\centering
\includegraphics[width=4cm,height=12cm,angle=-90]{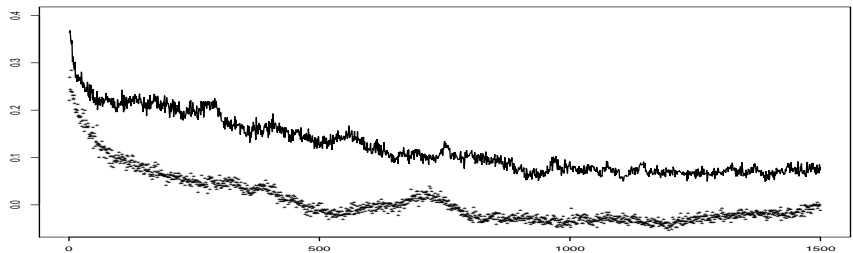}
\caption{The first 1500 values of the empirical ACF of the S+P(500)
  data for the first (lines) and second (*) halves of the data.}
\label{fig:acf2}
\end{figure}

Whether quantitative or qualitative comparisons are made there is one
fundamental problem with the S+P 500 data sets, namely that there
are no independent comparable data sets. This means that it is is
difficult to judge the variability of such data sets. As an example
some of the autocorrelations functions generated by the GARCH(1,1)
process shown in  Figure~\ref{fig:acf} may be judged as being too
extreme to be credible for long range financial
data. Figure~\ref{fig:acf2} shows the first 1500 lags for
the first half of the data points (lines) and the same for the second
half (*). This suggests that the variability of the
autocorrelation functions for the absolute returns can indeed be
quite large even for very long data sets.

\section{Stylized facts and their quantification}
In the context of financial data a list of eleven stylized facts is
given in \cite{CON01}. The ones to be considered in this paper are
{\bf 1. Absence of autocorrelations}, {\bf 2. Heavy tails}, {\bf
  3. Gain/loss asymmetry}, {\bf
  6. Volatility clustering}, {\bf 7. Conditional heavy tails}, {\bf
  8. Slow decay of autocorrelation in absolute terms} and {\bf
  10. Leverage effect}. These stylized
facts are exhibited by most to all of the data sets we consider.
As stated above we shall be concerned with the ability of a model to
reproduce these stylized facts in a quantitative sense for a given
empirical time series. In some cases the quantification is
straightforward, in other cases, and in particular for volatility
clustering, there is no obvious manner in which this stylized fact can
be quantified.

\subsection{Absence of autocorrelations}
The autocorrelations are not absent but small. The question is how is
small to be defined. The value of the first lag for the signs of the
S+P 500 data is 0.0577 which is certainly statistically  significant
but may not be practically relevant. The course taken in this paper is
to reproduce the value of the first lag of the ACF but the software
allows the user to produce other values.

Let $eac1$ denote the value of the first lag of the ACF of the signs
of the data and $sac1(i), i=1,\ldots, nsim$ be the simulated
values. The $p$-value of $eac1$ is defined by
\begin{equation}
p=\min(p_1,p_2)
\end{equation}
where
\begin{equation}
p_1=\#\{i: sac1(i) \le eac1\}/nsim\,\,\text{and}\,\, p_2=\#\{i: sac1(i)\ge eac1\}/nsim\,.
\end{equation}
The $p$-value is a measure of the extent to which the empirical values
can be reproduced in the simulations. It is seen that $0\le p\le
1/2$. This definition of a $p$-value will apply to any statistic
whose value may be too small or too large. In some cases,
 only values which are too large are of interest and a
one-sided definition will be used.

\subsection{Heavy tails} \label{sec:heavy_tails}
A standard way of quantifying heavy tails is to use the kurtosis. The
kurtosis is however extremely sensitive to outliers. The S+P 500 data
give an example of this. The largest absolute value of the data is
-0.229 and if this single value is removed the kurtosis drops from
21.51 to 15.37. Because of this extreme sensitivity the kurtosis will
not be considered any further. Instead the following measure will be
used. Denote the ordered absolute values of the returns
normalized by their median by $eaq$ and by $aq$ the corresponding
values for the normal distribution whereby the quantile values
$\text{qnorm}(i/(n+1))$ are used. The measure of the heaviness of the
tails is taken to be the mean of the difference $eaq-aq$. For normal
data the value is close to zero. For $n=23000$ the values for data
with a $t$-distribution with 2 and 3 degrees of freedom the values are
0.451 and 0.226 respectively where the quantiles
$\text{qt}(i/23001,\nu), \nu=2,3$ were used. The value for the S+P 500 is 0.316.

\subsection{Gain/loss asymmetry} \label{sec:sign}
\begin{figure}[!]
\begin{center}
\includegraphics[width=4cm,height=12cm,angle=-90]{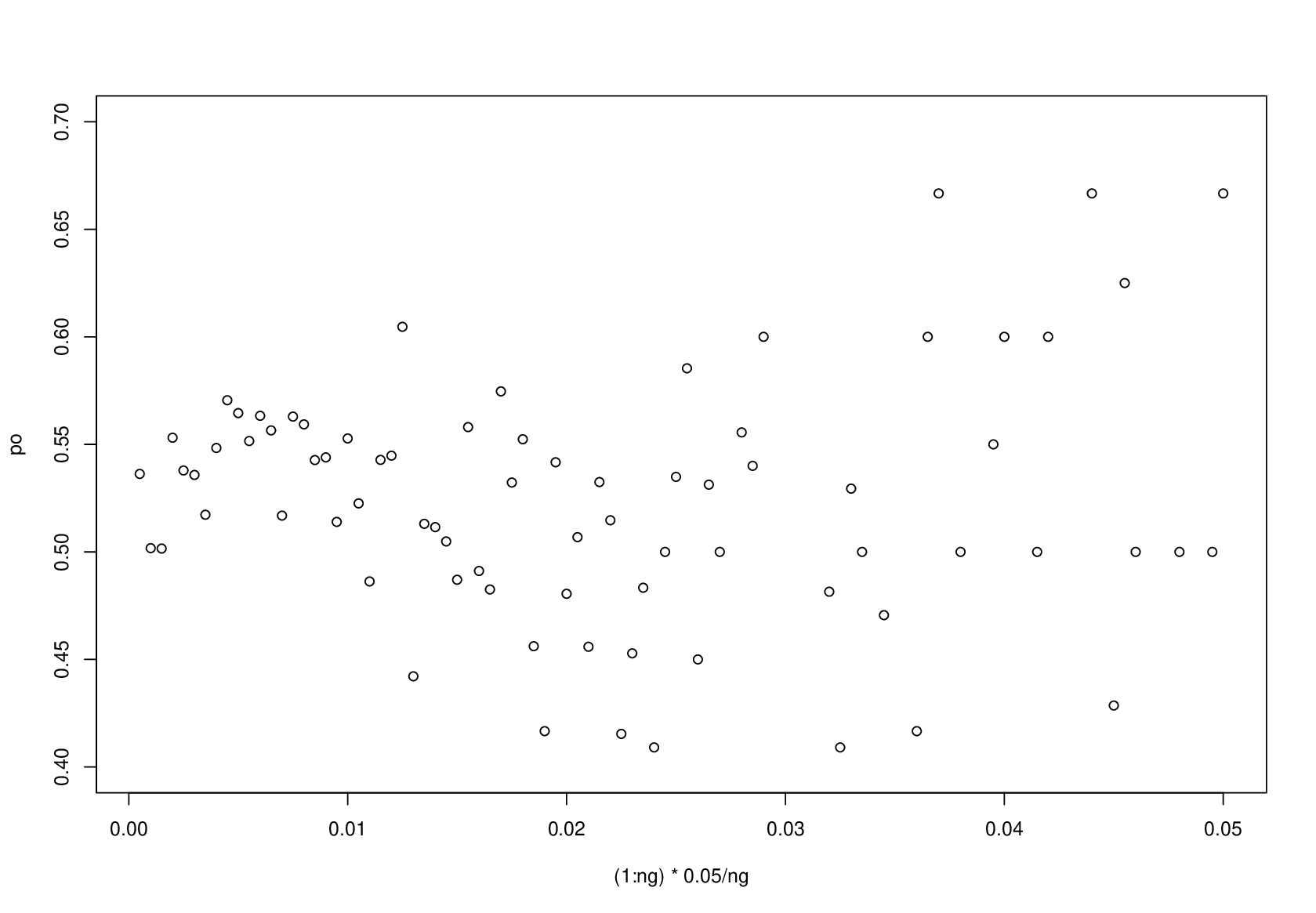}
\includegraphics[width=4cm,height=12cm,angle=-90]{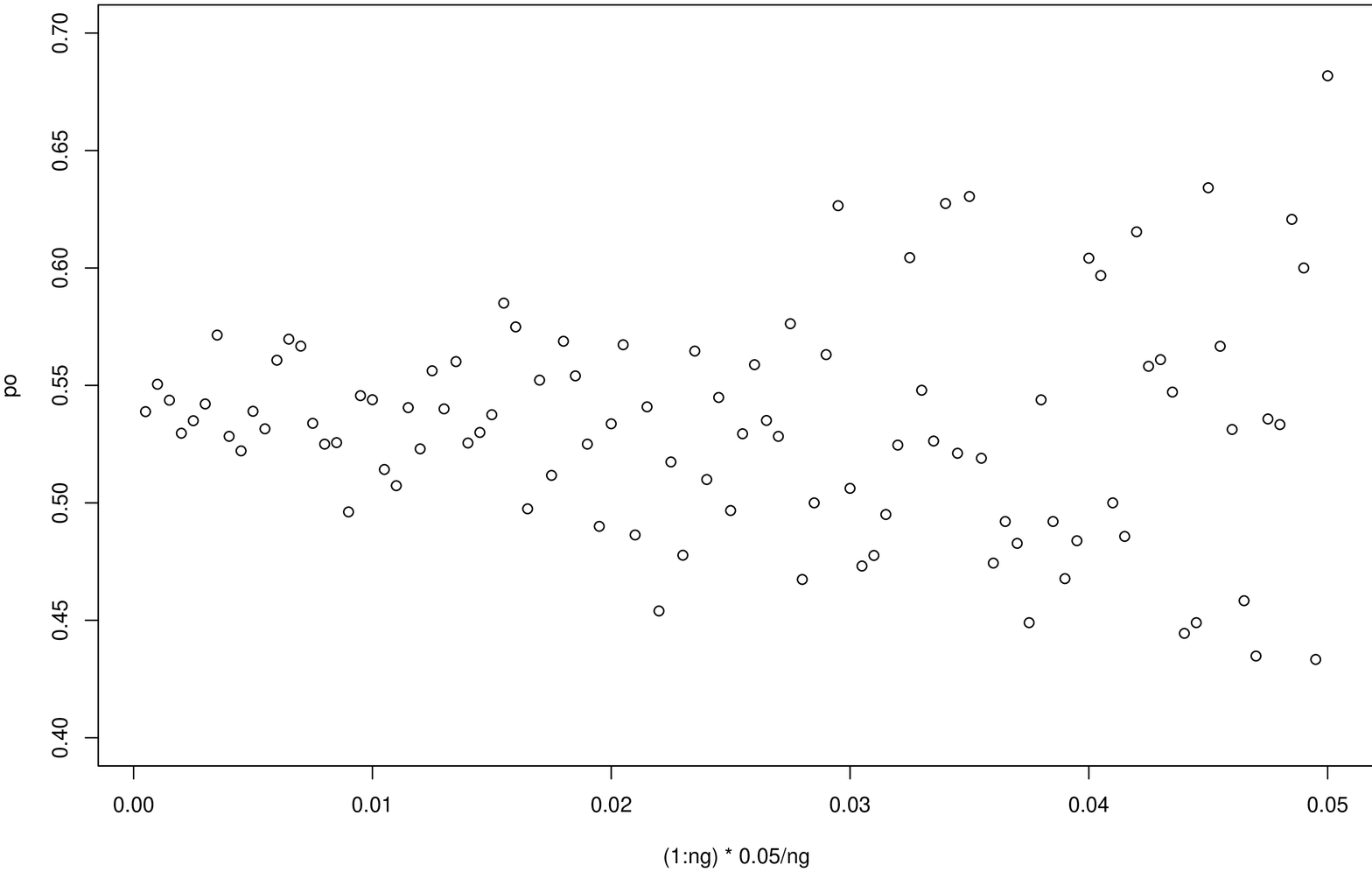}
\includegraphics[width=4cm,height=12cm,angle=-90]{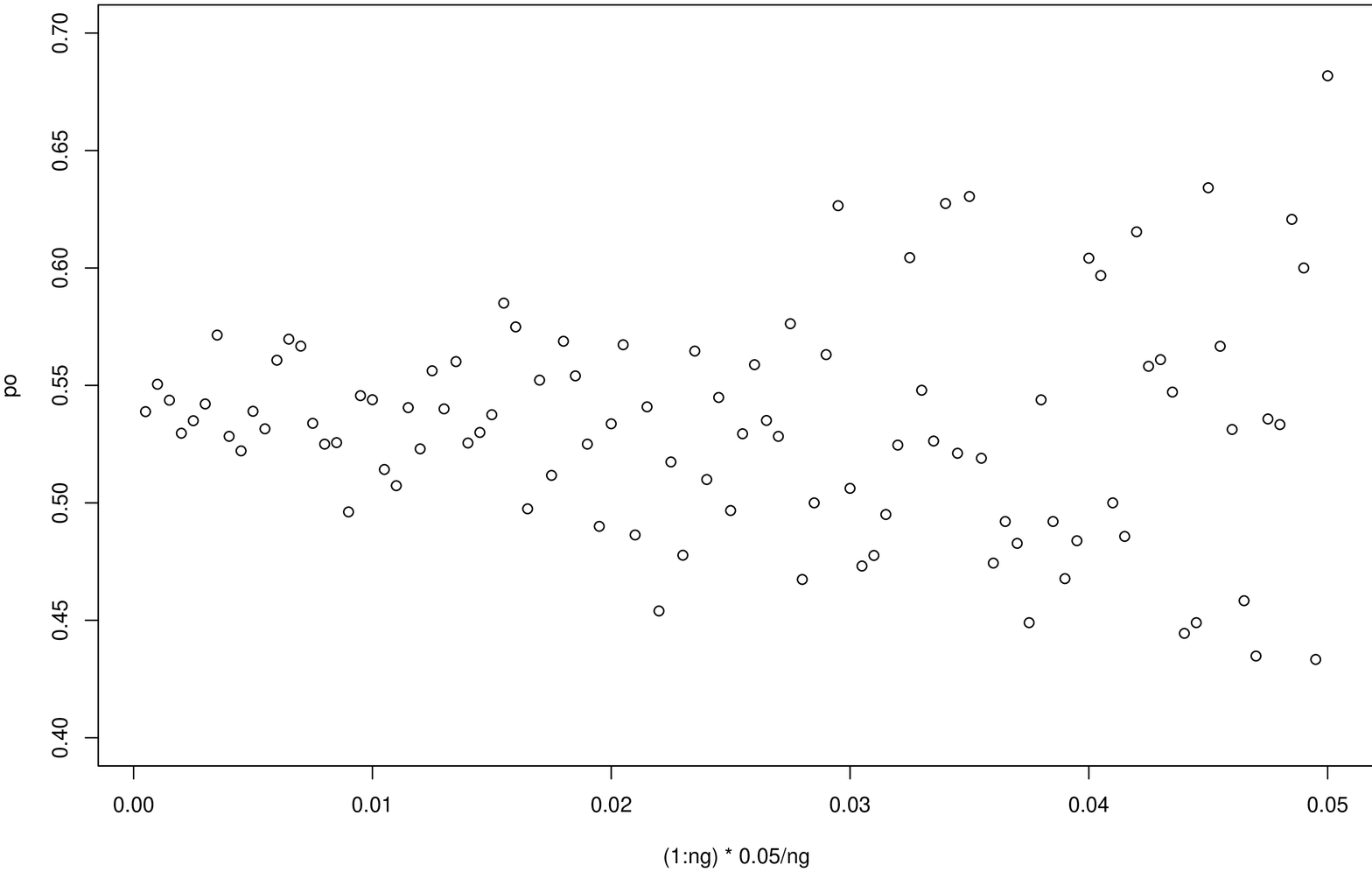}
\end{center}
\caption{Top panel: proportion of positive returns as a  function of
  the absolute return for the S+P 500 data. Centre panel: the same for
  the DAX data. Bottom panel: the same for Heidelberger Zement\label{fig:sign12}}
\end{figure}

The top panel of  Figure~\ref{fig:sign12} shows the relative
frequency of a positive  return as a function of the absolute size of
the return for the S+P 500 data. The centre panel shows the same for
the DAX data and the bottom panel the same for Heidelberger
Zement, the latter is  based on the 9427 days where the return was not
zero. The correlations are -0.480, -0.140 and 0.354 respectively. The
plots are calculated on the basis of the 0.02-0.98 quantiles of the
absolute returns. The S+P 500 and the DAX data are consistent with the
remark in  \cite{CON01} that `one observes large drawdowns in stock
prices and stock index values but not equally large upward
movements' and also `most measures of volatility of an asset are
negatively correlated with the returns of that asset'. The
Heidelberger Zement data shows that this is not always the case.

Other things being equal, which they may not be, a dependency between
the absolute size of a return and its sign will induce an asymmetry in
the distribution of the returns. As a measure of symmetry we use the
Kuiper distance
\[d_{\text{ku}}(\ep_{n_+}^+,\ep_{n_-}^-)\]
 between the distributions of the positive and negative returns. This
 may be seen as a variant of the two-sample Kolmogorov-Smirnov
 test. The Kuiper values for the S+P 500, the DAX and Heidelberger
 Zement data sets are 0.0412, 0.0290 and 0.0342 with (asymptotic)
 $p$-values 0.000, 0.060 and 0.0810 respectively.

\subsection{Volatility clustering}  \label{sec:vol_clust}
\begin{figure}[!h]
\centering
\includegraphics[width=4cm,height=12cm,angle=-90]{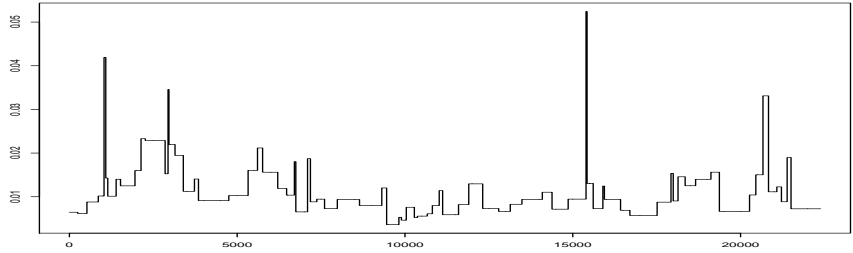}
\caption{The 76 intervals of constant volatility of the the S+P 500 data.}
\label{fig:stpr_garch_vol}
\end{figure}

The quantification of volatility clustering is the most difficult
stylized fact to quantify in spite of its visual clarity. The
quantification we shall use is based on \cite{DAVHOEKRA12}. The
basic model is
\begin{equation} \label{equ:basic_model}
R_t=\Sigma_tZ_t
\end{equation}
where $Z$ is standard Gaussian noise. From this it follows
\begin{equation}\label{equ:chi2}
\sum_{t=i}^j\frac{R_t^2}{\Sigma_t^2}\stackrel{D}{=}\chi^2_{j-i+1}
\end{equation}
and hence
\begin{equation}\label{equ:chi2_inequ}
\text{qchisq}((1-\alpha)/2,j-i+1)\le \sum_{t=i}^j\frac{R_t^2}{\Sigma_t^2}\le
 \text{qchisq}((1+\alpha)/2,j-i+1)
\end{equation}
with probability $\alpha$. These latter inequalities form the basis of
\cite{DAVHOEKRA12} where they are extended from one fixed interval
$[i,j]$ to a family of intervals ${\mathcal F}$ which form a local
multiscale scheme. In this case the $\alpha$ of (\ref{equ:chi2_inequ})
must be replaced by $\alpha_n$.  The goal is to determine a piecewise
constant volatility $\Sigma_t$ which satisfies the inequalities
(\ref{equ:chi2_inequ}) for all $[i,j]\in {\mathcal F}$. This problem
is ill-posed. It is regularized by requiring that $\Sigma_t$ minimizes
the number of intervals of constancy subject to the bounds and to the
values of $\Sigma_t$ on an interval of constancy being the empirical
volatility on that interval. Finally $\alpha_n$ is chosen by
specifying an $\alpha$ and requiring that the solution is one single
interval with probability $\alpha$ if the data are standard Gaussian
white noise: see \cite{DAVHOEKRA12} for the details.  For $\alpha=0.9$
and $n=22381$ the value of $\alpha_n$ is 0.9999993. For the S+P 500
data there are 76 intervals of constancy. They are
shown in Figure~\ref{fig:stpr_garch_vol}.

The $Z_t$ in (\ref{equ:basic_model}) can be replaced by other forms of
white noise, for example a $t$-random variable with a given number of
degrees of freedom. This gives a better fit but comes at the cost of
an increase in computational complexity (see Chapter~8 of \cite{DAV14}).

\begin{figure}[!h]
\centering
\includegraphics[width=4cm,height=12cm,angle=-90]{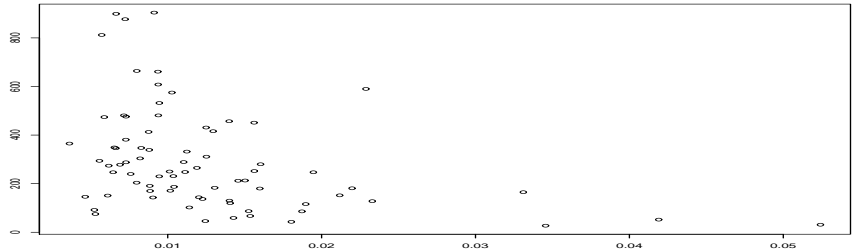}
\caption{Sojourn times as a function of volatility for the S+P 500.}
\label{fig:stpr_soj}
\end{figure}

In this paper the number of clusters will be used as a measure of the
degree of clustering or the volatility of the volatility. There are
other possibilities such as the sizes of the clusters
(Figure~\ref{fig:stpr_soj} shows the sojourn times plotted against the
volatility for the 76 intervals) but this and other measures will not
be considered further.

\subsection{Conditional heavy tails}
The claim in \cite{CON01} is that `even after correcting returns for
volatility clustering .... the residual time series still exhibit
heavy tails'.  This as stated is not sufficiently precise to enable a
numerical expression. If the model (\ref{equ:basic_model}) is used
with $Z$ standard Gaussian white noise and the volatility determined as
described in Section~\ref{sec:vol_clust} with $\alpha=0.9$ then the
residual times series has a kurtosis of 4.50 as against 3 for the
standard normal distribution. If the $Z$ in (\ref{equ:basic_model}) is
taken to have a $t$--distribution with 5 degrees of freedom and again
$\alpha=0.9$ then the residual times series has a kurtosis of 6.087 as
against 6 for the $t_5$-distribution. The matter will be discussed no
further.

\subsection{Slow decay of autocorrelation of absolute returns}
As already mentioned the upper panel of Figure~\ref{fig:acf} shows the
autocorrelation function of the absolute return of the S+P 500 for the
first 1500 lags (black) and the mean ACF based on 100 simulations of
the GARCH(1,1) model using the maximum likelihood parameters
(grey). The slow decay of the ACF for the  S+P 500 data is apparent.

As a measure of closeness of two autocorrelation  functions $a_1$ and
$a_2$ over $\ell$ lags we take the average absolute difference
\begin{equation}
d_{\text{acf}}(a_1,a_2)=\frac{1}{\ell}\sum_{i=1}^{\ell} \vert a_1(i)-a_2(i)\vert \,.
\end{equation}

The $p$-value for the autocorrelation function of a data set based on a
model is defined as follows. Let $a$ denote the mean ACF based on the
model. This can be obtained from simulations. In a second set of
simulations the distribution of $d_{\text{acf}}(A,a)$ can be
determined where $A$ denotes a random ACF based on the model. Given
this the $p$-value of  $d_{\text{acf}}(ea,a)$ can be obtained relative
to the distribution of  $d_{\text{acf}}(A,a)$ where $ea$ denotes the
ACF of the data. For the S+P 500 data with $\ell=1500$
$d_{\text{acf}}(ea,a)=0.0522$ with  $p$-value 0.061. The
0.95-quantile is 0.0571, the mean 0.0160 and the standard deviation
0.0224.

In \cite{TERZHA11} the value of the first lag of the ACF of the
absolute returns was considered. This too will be included in the
features to be reproduced.

\subsection{The end return} \label{sec:stylized5}
The end return is just the end value of the stock or index given a starting
value of one. We shall require that this is adequately reflected by the
model. Many models modify the basic model (\ref{equ:basic_model}) by
including an additive form for the drift
\begin{equation} \label{equ:drift}
R_t=\mu_t+\Sigma_tZ_t
\end{equation}
where the $Z_t$ are assumed to have zero mean. In such a model the
final return will depend on $\mu_t$. To simulate data some stochastic
assumptions must be made for $\mu_t$ and it is not clear how to do
this. We prefer to keep to the basic model (\ref{equ:basic_model}) but
to let the sign of $R_t$ depend on its absolute magnitude as in
Section~\ref{sec:sign}. It turns out that this is sufficient to
successfully reproduce the end return. We point out that this can also
be done for the GARCH(1,1) process without disturbing the generating
scheme.

\subsection{Absolute moments of the returns}\label{sec:stylized6}
Although they are not classified as stylized facts we shall require
that the  first and second absolute moments of the
daily returns are adequately reflected by the model.

\subsection{Quantiles and distribution of returns}
Finally we consider two measures of the distribution of the
returns. Denote by $eqm$, $rqm$ and $qm$ the order statistics of the
data, of a random simulation and the mean of the simulations
respectively. The mean absolute deviation of a random simulation is
\begin{equation} \label{equ:abs_dev_quant}
\frac{1}{n}\sum_{i=1}^n\vert rqm(i)-qm(i)\vert
\end{equation}
from which a one-sided p-value for the empirical deviation
\[\frac{1}{n}\sum_{i=1}^n\vert eqm(i)-qm(i)\vert\]
can be obtained.

The same applies for the Kuiper distances. With the obvious notation
the simulated Kuiper distances are $d_{\text{ku}}(\ep_{rqm},\ep_{qm})$
 from which again a one-sided p-value can be obtained for the
 empirical distance $d_{\text{ku}}(\ep_{eqm},\ep_{qm})$.

\subsection{List of quantified features to be reproduced} \label{sec:list}
In all there are eleven quantified features which are to be reproduced
by the simulations. The degree to which this is accomplished will be
measured by either a one-sided or a two-sided p-value as appropriate.

\begin{enumerate}
\item First autocorrelation of the signs of the returns
\item Heavy tails
\item Symmetry/asymmetry of returns
\item Volatility clustering - number of intervals of constant
  volatility
\item Slow decay of the ACF of absolute returns
\item Value of first lag of the ACF of absolute returns
\item Final return
\item Mean of absolute returns
\item Mean of squared returns
\item Quantiles of returns
\item Kuiper distance of returns
\end{enumerate}

\section{Modelling the data} \label{sec:modelling}
In \cite{STUA03} the GARCH(1,1) model is explicitly used as an example
of a stationary parametric model. In the literature however it seems
to be generally accepted that the S+P 500 cannot be satisfactorily
modelled using this or any other stationary parametric model, see for
example \cite{MIKSTA04} and  \cite{GRASTA05}. If this is so then
alternative forms of modelling must be used. Possibilities  are to use
locally stationary models \cite{DAHRAO06}, segment the data and to use
stationary models in each segment (\cite{GRASTA05}), to use a
semi-parametric approach (\cite{NETDARTES12}, \cite{AMATER14})
or a non-parametric approach (\cite{MIKSTA03}, \cite{DAVHOEKRA12},
\cite{NETDARTES12}) whereby the boundaries between the three
approaches are somewhat fluid. There are also more ambitious models
which attempt to reproduce some stylized facts at least qualitatively
by modelling the activities of the agents (see for example
\cite{HOM02}, \cite{LUSCH05} and \cite{CON07}). It seems to be
difficult to adapt these to a quantitative reproduction of a particular
stock.

Whether a time series is regarded as stationary, that is, it can be
satisfactorily modelled by a stationary process, depends on the time
horizon.  Data which may not look stationary on a short horizon may be
part of a data set which looks stationary on a longer horizon. Any
finite data set can be embedded in a stationary process as
follows. The data are extended periodically in both directions and
then the origin chosen using a uniformly distributed random variable
over the original data set. This may not be the best way of claiming
that the data are part of a stationary process but it does show that
the question of stationarity is ill-posed.

The basic model is (\ref{equ:basic_model}). The modelling will be
done in two steps, firstly modelling the volatility process $\Sigma_t$
and then the `residual' process $Z_t$.

\subsection{Modelling the volatility $\Sigma_t$}
\begin{figure}[b]
\begin{center}
\includegraphics[width=4cm,height=12cm,angle=-90]{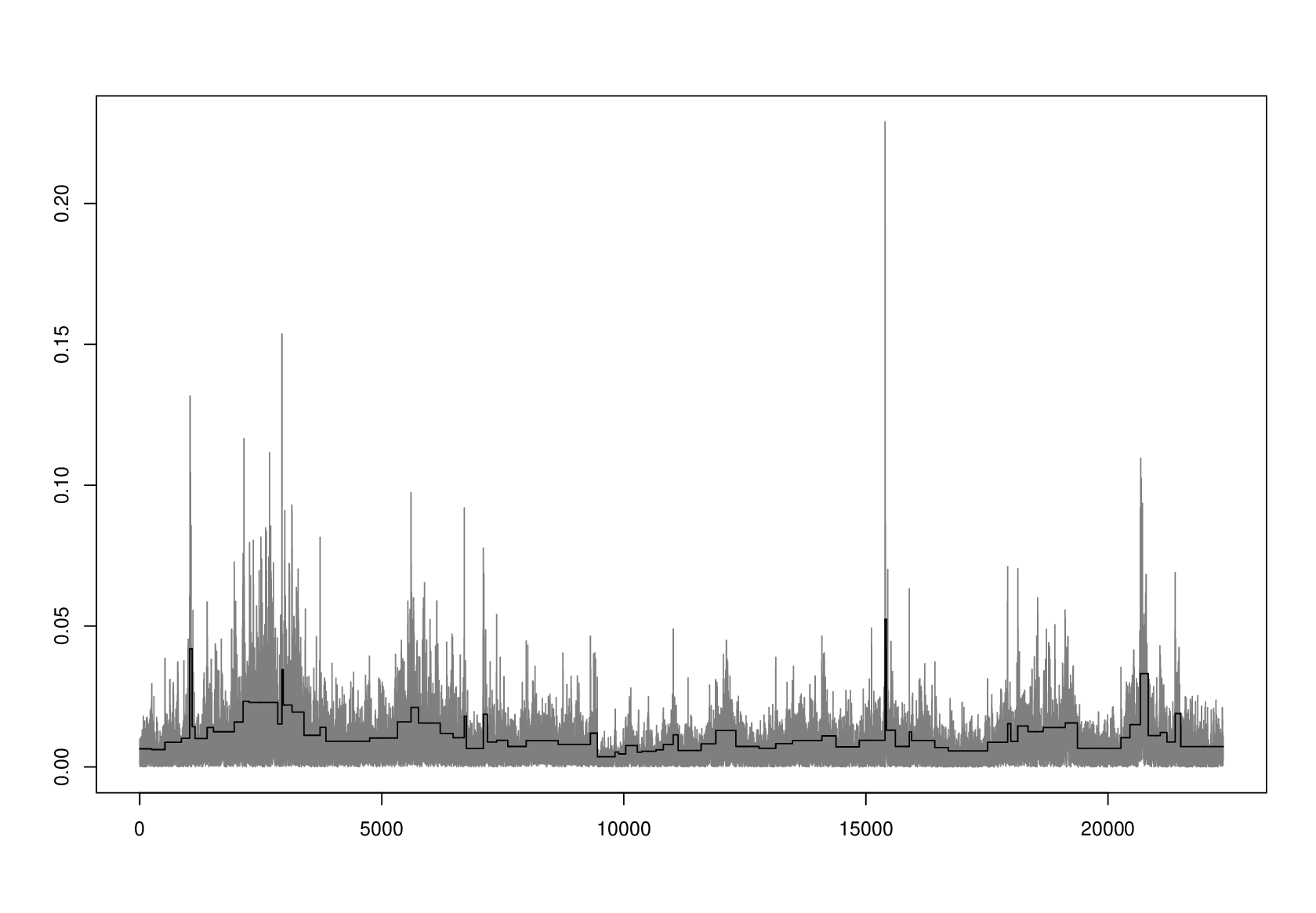}
\includegraphics[width=4cm,height=12cm,angle=-90]{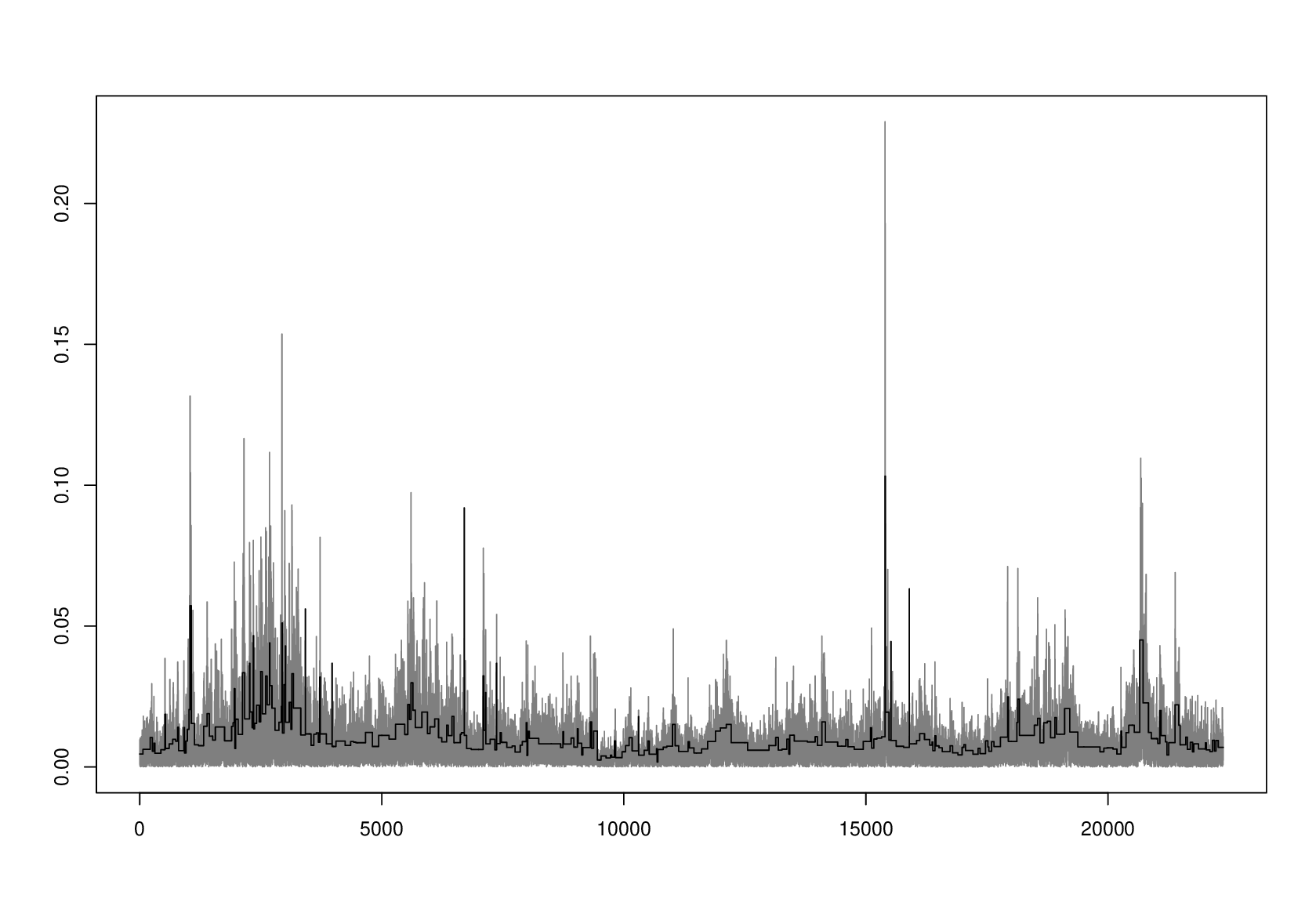}
\end{center}
\caption{The  absolute daily returns of the Standard and Poor's index
  together with a piecewise constant volatility: upper panel with
  $\alpha_n=0.9999993$ and 76 interval; lower panel with
  $\alpha_n=0.998$ and 283 intervals. \label{fig:stpr12}}
\end{figure}

The upper panel of Figure~\ref{fig:stpr12} shows the absolute returns of
the Standard and Poor's data together with the piecewise constant
approximation of the volatility (\cite{DAVHOEKRA12}) using the value
$\alpha_n=0.9999993$. The construction of the piecewise constant
volatility is a form of smoothing and as such small local variations
in volatility will be subsumed in a larger interval of constant
volatility. Choosing a smaller value of $\alpha_n$ allows the
reconstruction of smaller local changes. The lower panel of
Figure~\ref{fig:stpr12} shows the absolute returns of the Standard and
Poor's data together with the piecewise constant approximation of the
volatility with $\alpha_n=0.998$. There are 283 intervals of
constancy. The choice of $\alpha_n$ is the first screw
which can be tightened or slackened.

\begin{figure}[!]
\begin{center}
\includegraphics[width=4cm,height=12cm,angle=-90]{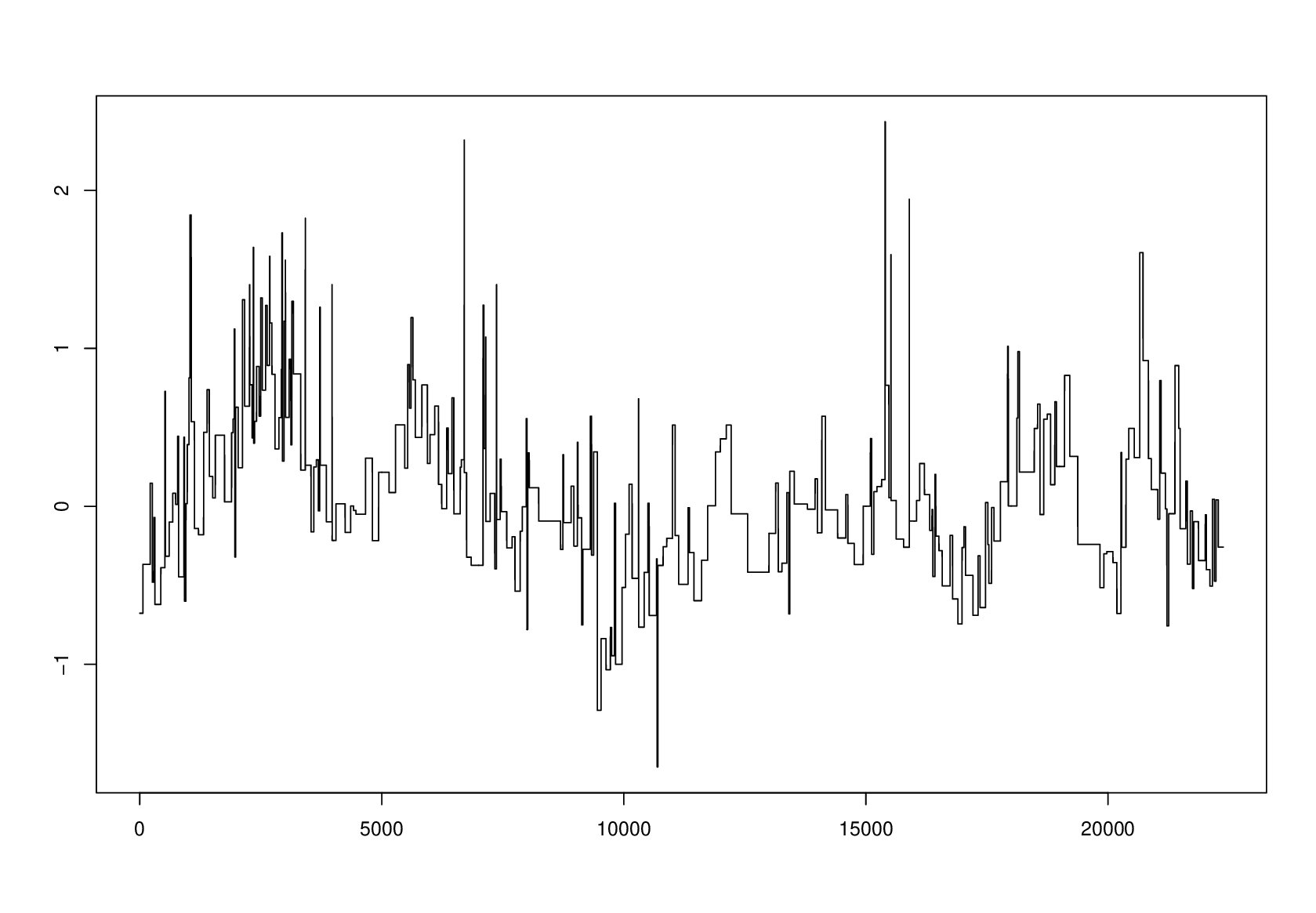}
\includegraphics[width=4cm,height=12cm,angle=-90]{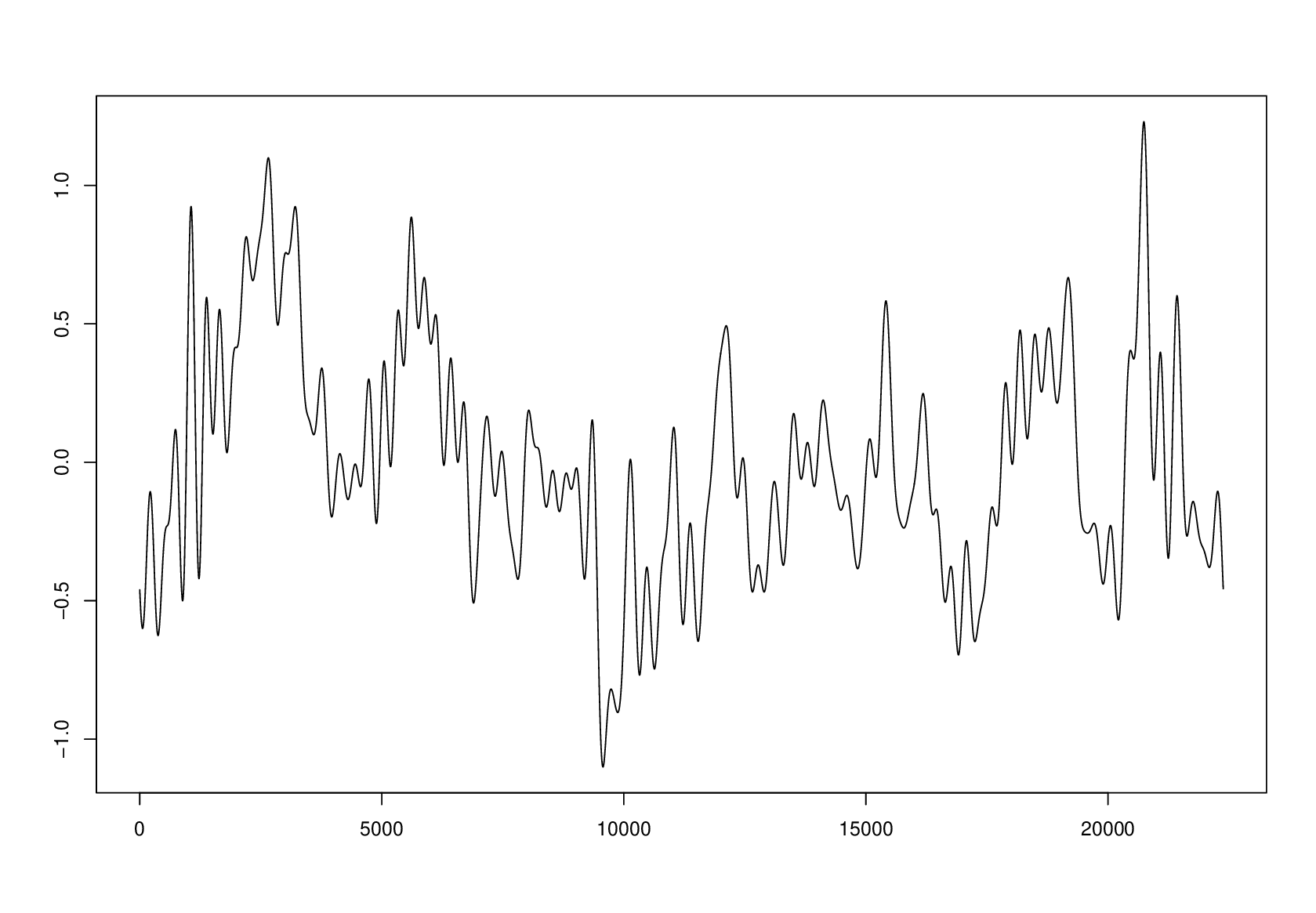}
\includegraphics[width=4cm,height=12cm,angle=-90]{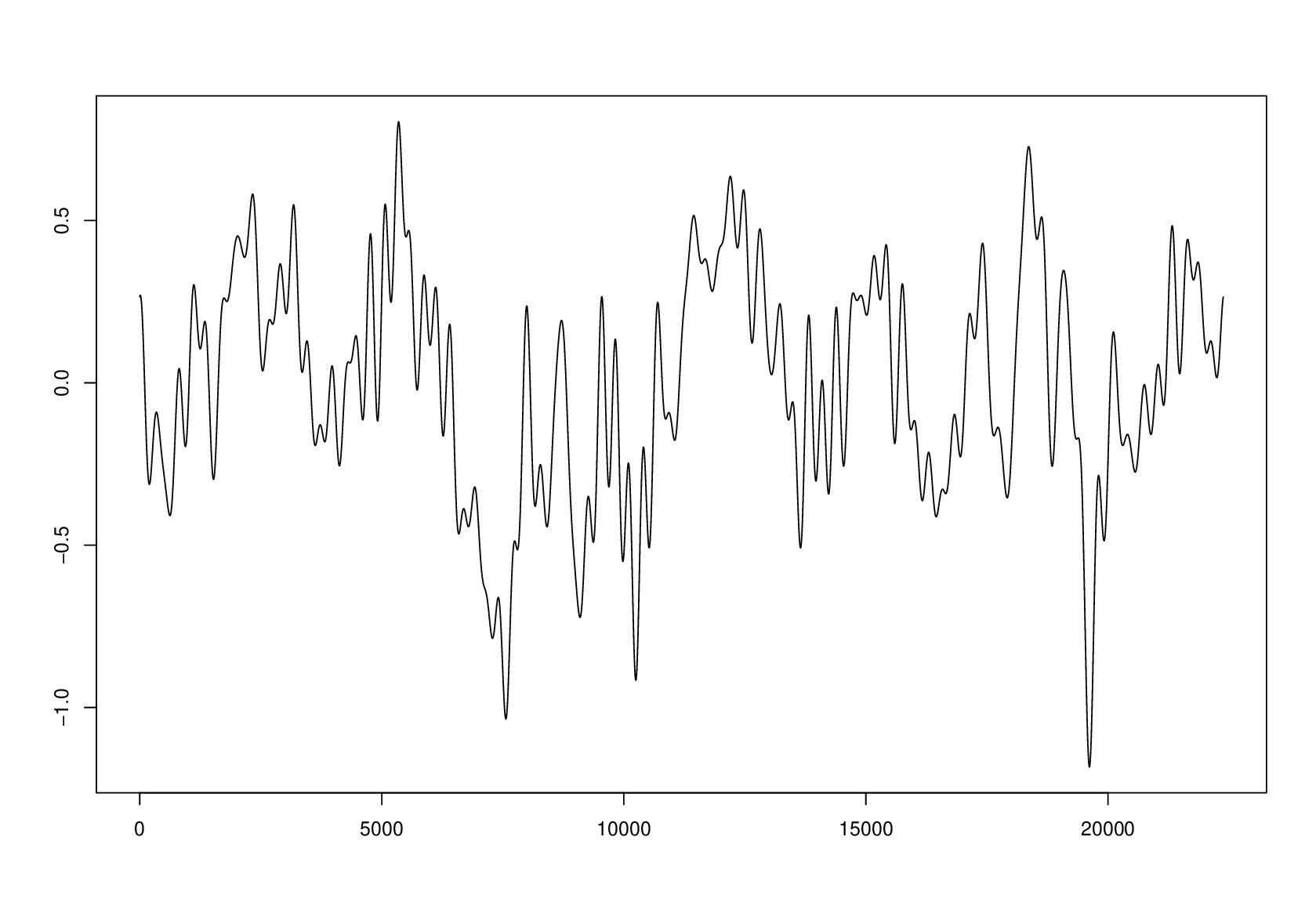}
\end{center}
\caption{Top panel: the logarithm of the piecewise constant volatilities of the
  lower panel of Figure~\ref{fig:stpr12}. Centre panel: an
  approximating trigonometric polynomial  accounting for 80\% of the
  variance. Bottom panel: a randomized version of the
  polynomial. \label{fig:stpr_lv}}
\end{figure}

In a first step the log-volatilities are centred at zero by
subtracting the mean. They are then approximated by a low order
trigonometric polynomial
\begin{equation} \label{equ:trig_poly0}
p_j(k)=a_j\sin(2\pi jk/n)+b_j\cos(2\pi j k/n), \quad k=1,\ldots,n
\end{equation}
where the coefficients $a_j$ and $b_j$ are determined by least
squares. The calculation can be made considerably faster by using the
Fast Fourier Transform. The number of polynomials is determined by a
further screw $pow$ which gives the proportion of the total variance of the
log-volatilities to be accounted for by the polynomial. The top panel
of Figure~\ref{fig:stpr_lv} shows the logarithm of the piecewise
constant volatilities centred at zero, the centre panel shows the
approximating polynomial with $pow=0.8$ which is composed of 97
polynomials of the form (\ref{equ:trig_poly0}). This may be regarded
as a low frequency approximation to the logarithms of the piecewise
constant volatility.

The polynomials of (\ref{equ:trig_poly0}) are randomized by
multiplying the coefficients $a_j$ and $b_j$ by standard independent
Gaussian random variables:
\begin{equation} \label{equ:trig_poly1}
Z_{1j}a_j\sin(2\pi jk/n)+Z_{2j}b_j\cos(2\pi j k/n), \quad k=1,\ldots,n\,.
\end{equation}
The bottom panel of  Figure~\ref{fig:stpr_lv} shows a randomized
version of the polynomial of the centre panel.

\begin{figure}[!]
\begin{center}
\includegraphics[width=4cm,height=12cm,angle=-90]{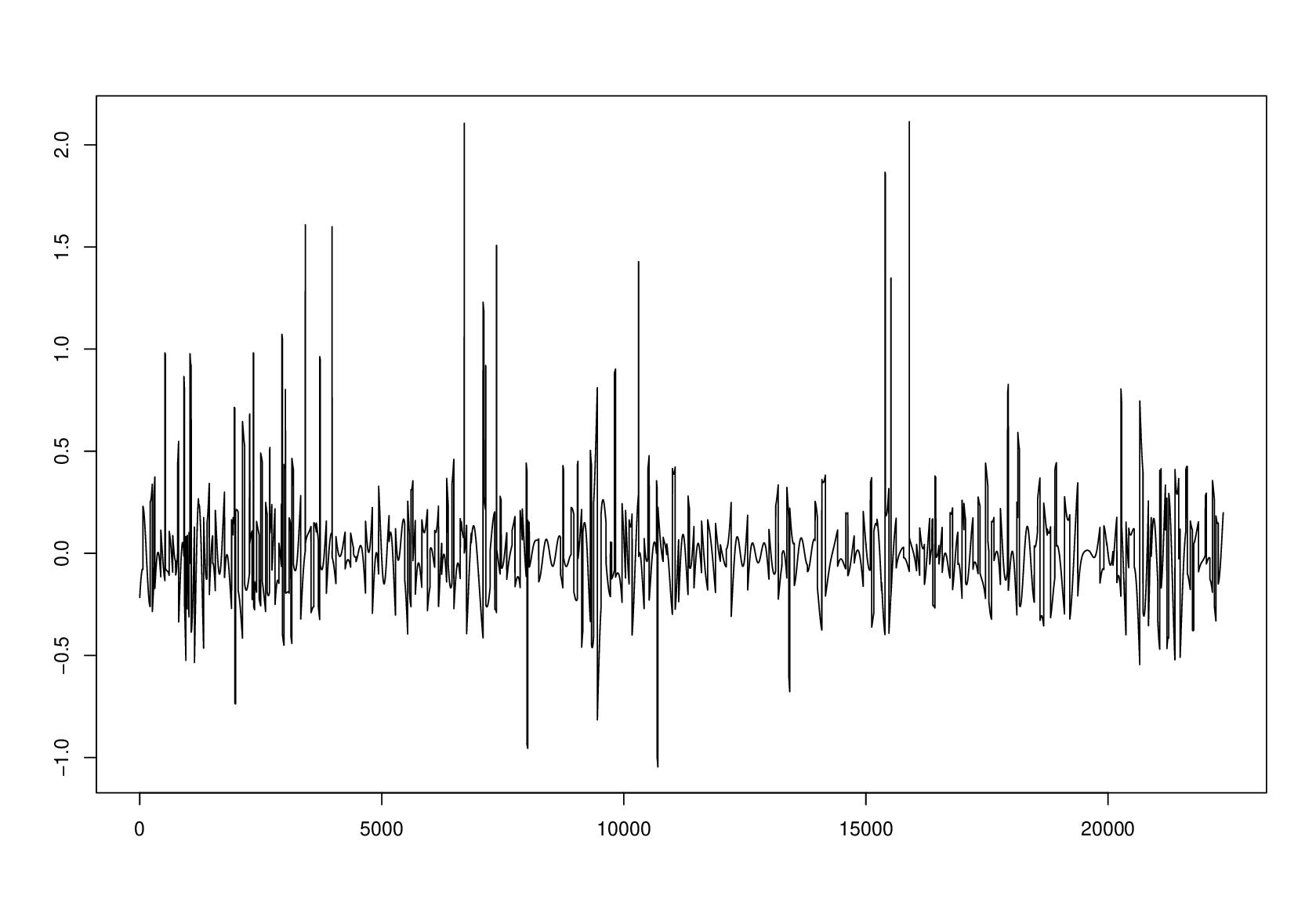}
\includegraphics[width=4cm,height=12cm,angle=-90]{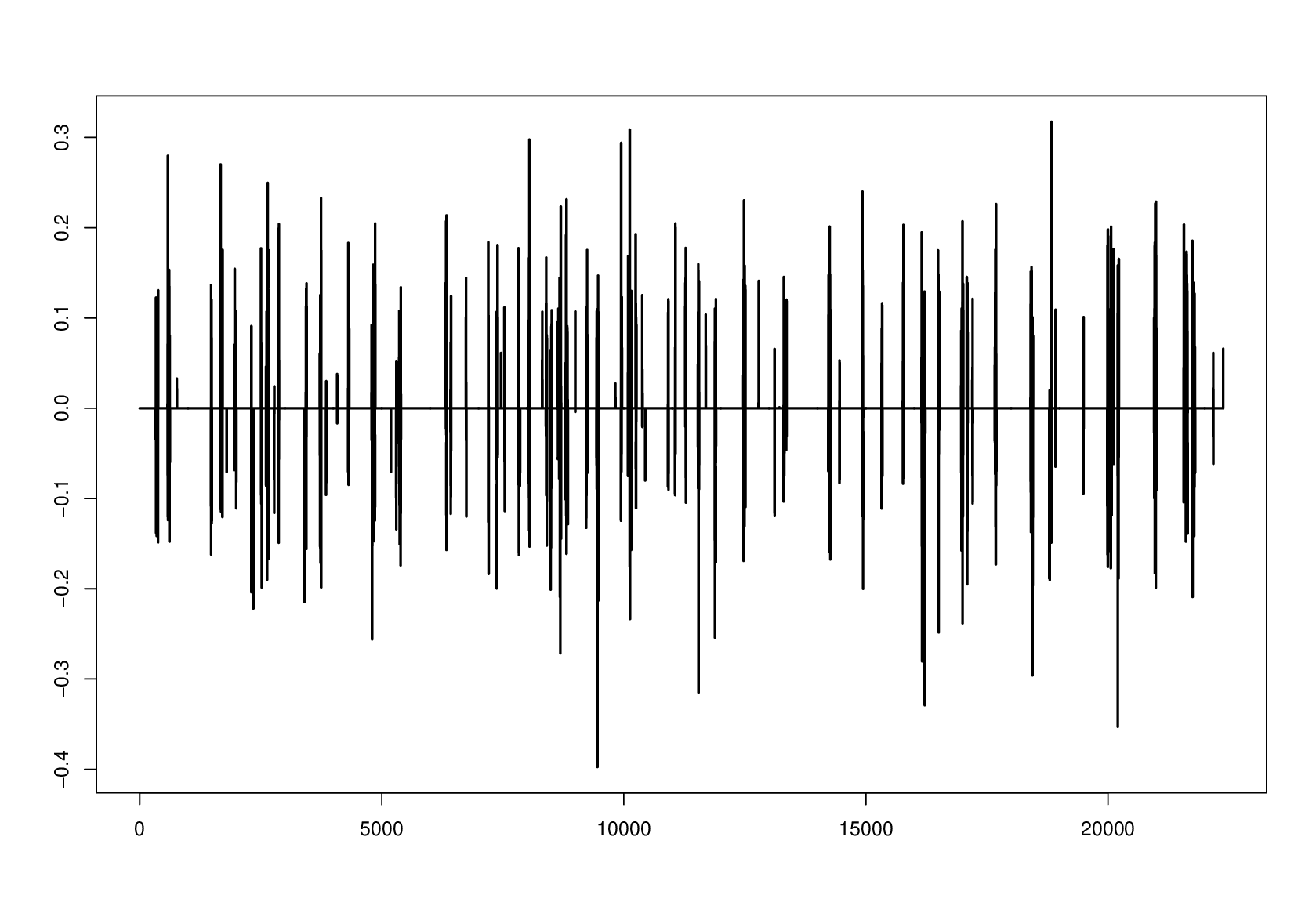}
\end{center}
\caption{Upper panel:The residuals of the log-volatility after removal
  of the low frequency approximation. Lower panel: a randomization of
  the residuals. \label{fig:stpr_lv_high}}
\end{figure}

\begin{figure}[!]
\begin{center}
\includegraphics[width=4cm,height=12cm,angle=-90]{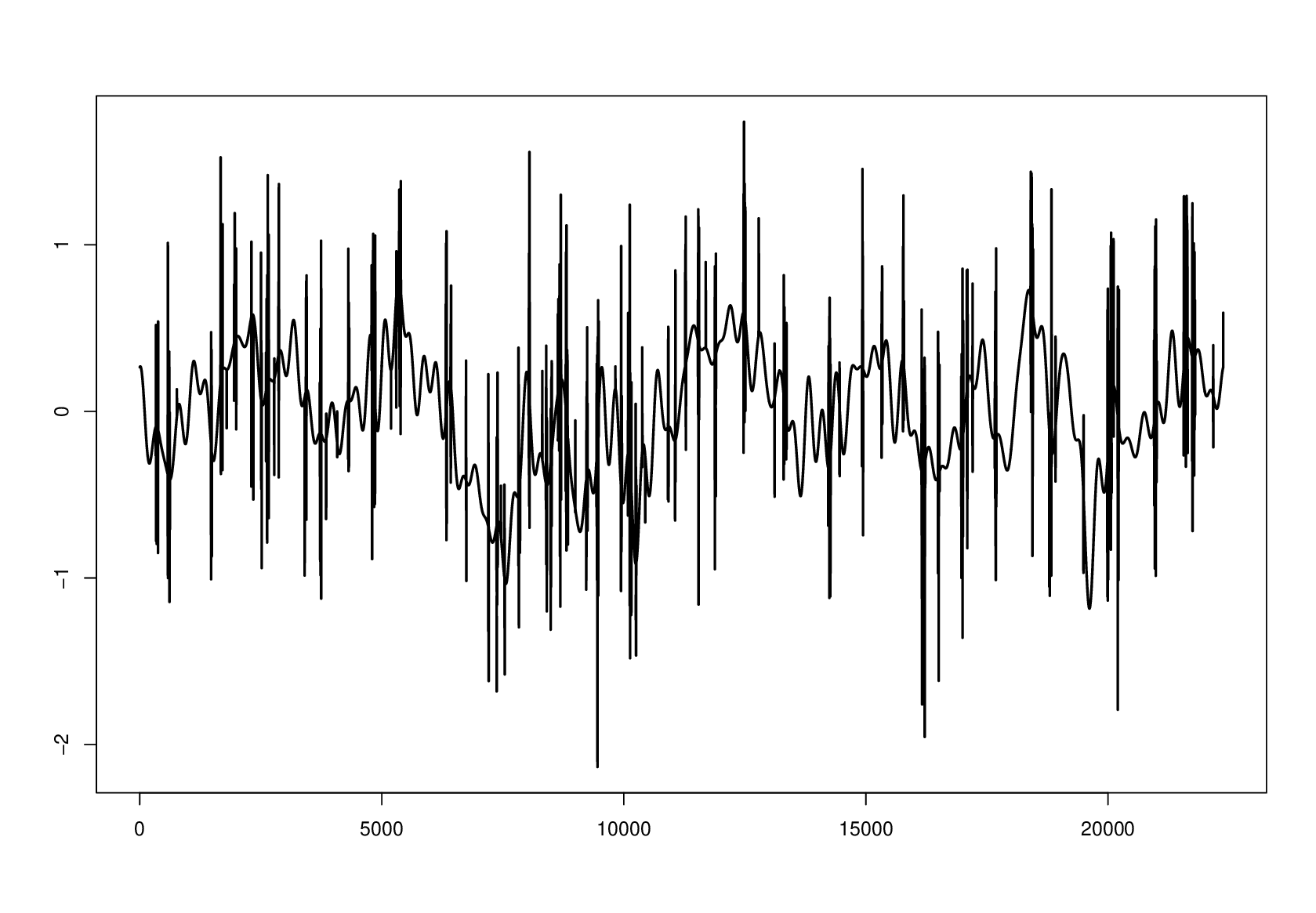}
\end{center}
\caption{A randomization of the S+P 500 volatility process. \label{fig:stpr_lv_ran_fin}}
\end{figure}

After removing the low frequency approximation the residuals form the
remaining high frequency log-volatility. They are shown in the upper
panel of  Figure~\ref{fig:stpr_lv_high}. It is not obvious how the
residuals can be modelled. In the following it will be done by
generating random intervals with lengths exponentially distributed
with alternating means $\lambda_1$ and $\lambda_2$. On the long intervals
corresponding to $\lambda_1$ the log-volatility will be modelled as
$N(0,\sigma_1^2)$ with $\sigma_1=0$ as the default value. On the short
intervals corresponding to $\lambda_2$ the log-volatility will be
modelled $\sigma_2T_{\nu}$ where $T_{\nu}$ is $t$-distributed with
$\nu$ degrees of freedom. The lower panel of
Figure~\ref{fig:stpr_lv_high} shows such a randomization with
$\lambda_1=200,\sigma_1=0,\lambda_2=20,\sigma_2=0.4,\nu=15$. Adding
the low and frequency components gives a randomization of the
volatility as shown in Figure~\ref{fig:stpr_lv_ran_fin}.

So far the smoothed log-volatility process has been centred at zero by
subtracting the mean $mlv=-4.769$. The problem now is to specify the
variability of the mean in the model. Some orientation can be obtained
by dividing the data into quarters and calculating the empirical mean
log-volatilities for each quarter. They are -4.451, -5.120, -4.807 and
-4.694. Based on this the mean will be modelled $mlv+\Delta$ where
$\Delta$ is uniformly distributed over an interval $[-\delta,\delta]$
with default value $\delta=0.2$. This concludes the modelling of the
process volatility process $\Delta_t$.

\subsection{Modelling the $R_t$} \label{sec:mod_R_t}
Let $\Sigma_t$ be the volatility process described in the last section
and ${\hat Z}_t$ be i.i.d. standard normal random variables. In a
first step put
\begin{equation} \label{equ:rho}
 {\tilde Z}_t=(\rho\vert {\hat Z}_{t-1}\vert+1){\hat Z}_t/
 \sqrt{1+2\rho\sqrt{2/\pi}+\rho^2} .
\end{equation}
The value of $\rho$ can be chosen so that the sixth item in the list
of Section~\ref{sec:list} can be adequately reproduced.

Given the ${\tilde Z}_t$ the absolute return is set to
\begin{equation} \label{equ:mod_R1}
\vert R_t\vert =\Sigma_t\vert Z_t\vert =\Sigma_t \vert {\tilde
  Z}_t\vert(1+\vert {\tilde Z}_t\vert)^{\eta}
\end{equation}
for some value of $\eta$ with default value zero.  A positive value of
$\eta$ makes the tails of $\vert Z_t\vert$ heavier than those of the
normal distribution, a negative value makes them lighter. It remains
to model the sign of the return.

As is evident form Figure~\ref{fig:sign12} the signs of the $R_t$ may
well depend on the value of $\vert R_t\vert$. This is taken into account
as follows. Denote the $i/\nu,i=1,\ldots,\nu$ quantiles of the
absolute returns of the data by $\text{eqa}(i)$ and the relative
frequency of the number of positive returns for  those returns $r_t$
with $\text{qa}(i-1) < \vert r_t\vert \le \text{qa}(i)$ by $p(i)$. The
default value of $\nu$ is $\nu =50$. Given a simulated value of the absolute
return $\vert R\vert$ with $\text{qa}(i-1) < \vert R\vert \le
\text{qa}(i)$ the actual return $R$ is taken to be positive with
probability $\gamma p(i)+(1-\gamma)/2$ where $\gamma,\, 0\le \gamma \le
1$. The parameter $\gamma$ is a further screw with default value $\gamma=1$.

Finally the first autocorrelation of the returns (first on the list of
Section~\ref{sec:list}) can be taken into account as follows. If the
empirical value of the autocorrelation is $eacf1$ then the final sign of
$R_t$ is determined as follows. Let $U_i, i=2,\ldots n$ be a sequence
of i.i.d. random variables uniformly distributed over $[0,1]$. If $U_i
>\vert eacf1\vert$ then the sign of $R_t$ is unchanged. If $U_i
<\vert eacf1\vert$ the the sign of $R_t$ is set equal to that of
$R_{t-1}$ if $eacf1>0$ and to the opposite sign of $R_{t-1}$ if
$eacf1<0$.


\section{The results of some simulations}
\begin{figure}[!h]
\centering
\includegraphics[width=4cm,height=12cm,angle=-90]{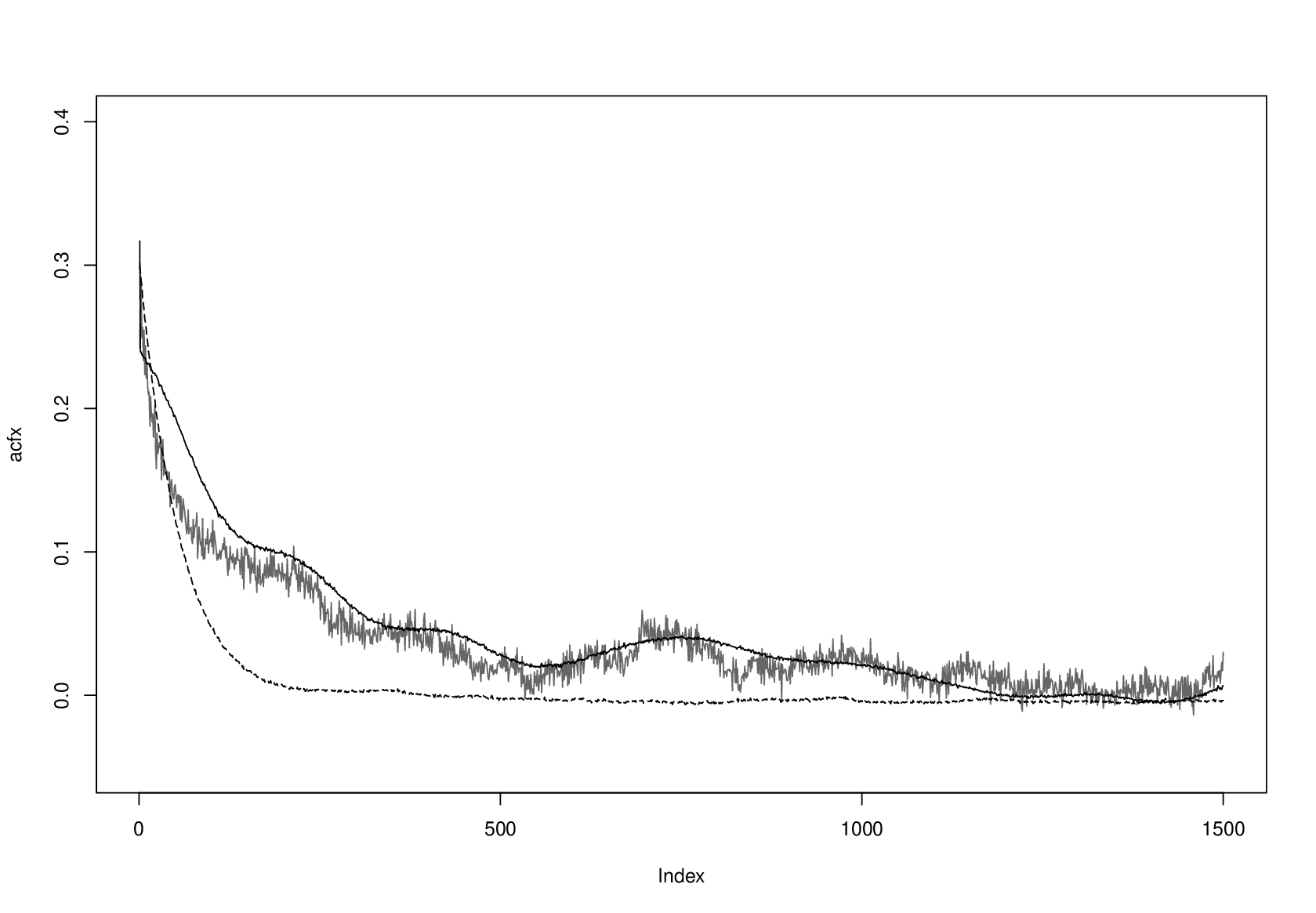}
\caption{The first 1500 values of the empirical ACF of the absolute
  values of the DAX data (grey), the mean ACF using the
  modelling of Section~\ref{sec:modelling} (black) and the mean ACF for
  the GARCH(1,1) modelling (dashed).}
\label{fig:dax_acf}
\end{figure}

The results for the S+P 500 and DAX data are given in
Table~\ref{tab:p-values}. They are given in terms of the p-values for
the 11 items of  Section~\ref{sec:list}. The starred items are
two-sided p-values with a maximum value of 0.5. The GARCH(1,1)
simulations have been modified to by altering the sign of the returns
as described in Section~\ref{sec:mod_R_t}. This has no effect on the
absolute values of the returns and consequently no further effect on
the GARCH(1,1) simulations. For the modelling described in
Section~\ref{sec:modelling} it proved possible in all cases to find
parameter values such that all p-values exceed 0.1. No attempt was
made to maximize the smallest p-values. The choice of parameter values
is not easy as most of them affect several features. This problem does
not occur for the GARCH(1,1) modelling. The best result for the
GARCH(1,1) model is when it is applied to the DAX data. There all but
two features have a p-value exceeding 0.1. The exceptions are heavy
tails $2^*$ where the GARCH(1,1) modelling results in tails which are
too light.  The worst failure is the inability to reproduce the slow
decay of the ACF of the absolute returns, feature
5. Figure~\ref{fig:dax_acf} shows the ACF of the DAX index (grey),
the mean ACF using the modelling described in
Section~\ref{sec:modelling} (black) and the mean for the GARCH(1,1)
modelling (dashed). 
\begin{table}
\begin{center}
\begin{tabular}{rccccccccccc}
&\multicolumn{11}{c}{Features as in Section~\ref{sec:list}}\\
&$1^*$&$2^*$&$3^*$&$4^*$&5&$6^*$&$7^*$&$8^*$&$9^*$&10&11\\
\hline
S+P 500&0.33&0.15&0.32&0.48&0.77&0.32&0.43&0.46&0.41&0.72&0.70\\
GARCH&0.32&0.06&0.17&0.12&0.05&0.30&0.48&0.00&0.07&0.03&0.00\\
\hline
DAX&0.33&0.16&0.24&0.46&0.88&0.50&0.48&0.29&0.30&0.54&0.51\\
GARCH&0.29&0.01&0.31&0.34&0.00&0.12&0.27&0.25&0.11&0.19&0.33\\
\hline
\end{tabular}
\caption{The p-values (based on 1000 simulations) for the 11 items on
  the list of stylized facts in  Section~\ref{sec:list} for the
  modelling of Section~\ref{sec:modelling}  and a modified GARCH(1,1)
  modelling.\label{tab:p-values}}
\end{center}
\end{table}

All thirty current members of the DAX were also modelled by both
methods. For the modelling described in Section~\ref{sec:modelling} it
was always possible to choose parameter values such that all 11
features had a p-value exceeding 0.1. The GARCH(1,1) modelling turned
out to be worse for these data sets than for the two indices S+P 500
and DAX. For all thirty firms the features 8-11 all had p-values of
zero. In the case of 8 and 9 the model underestimated the empirical
values in keep with the findings of \cite{STUA03} for a section of the
S+P 500 data. They were also underestimated for the S+P 500 data but
less severely. The DAX data are exceptional in this respect, the
empirical values were overestimated.

\bibliographystyle{apalike}
\bibliography{literature}
\end{document}